\documentclass[reprint,amsmath,amssymb,aps,
superscriptaddress,
]{revtex4-1}
\usepackage{CJK}
\usepackage{graphicx}
\usepackage{color}
\usepackage{dcolumn} 
\usepackage{bm}      
\usepackage{amsfonts}
\usepackage[bookmarks=false,colorlinks,citecolor=red]{hyperref}
\usepackage{caption}

\include{MyCommand}

\newif\ifshowcomments\showcommentstrue

\usepackage{changes}

\begin{document}

\title{Metal-Insulator and Magnetic Phase Diagram of Ca$_2$RuO$_4$\\ from Auxiliary Field Quantum Monte Carlo\\ and Dynamical Mean Field Theory}

\author{Hongxia Hao}
\affiliation{Department of Chemistry, Brown University, Providence, RI, 02912}
\author{Antoine Georges}
\affiliation{Coll{\`e}ge de France, 11 place Marcelin Berthelot, 75005 Paris, France}
\affiliation{Center for Computational Quantum Physics, Flatiron Institute, New York, NY 10010 USA}
\affiliation{CPHT, CNRS, {\'E}cole Polytechnique, IP Paris, F-91128 Palaiseau, France}
\affiliation{DQMP, Universit{\'e} de Gen{\`e}ve, 24 quai Ernest Ansermet, CH-1211 Gen{\`e}ve, Suisse}
\author{Andrew J.~Millis}
\affiliation{Center for Computational Quantum Physics, Flatiron Institute, New York, NY 10010 USA}
\affiliation{Department of Physics, Columbia University, New York, NY, 10027}
\author{Brenda Rubenstein}
\affiliation{Department of Chemistry, Brown University, Providence, RI, 02912}
\author{Qiang Han}
\affiliation{Department of Physics, Columbia University, New York, NY, 10027}
\author{Hao Shi}
\email{Hao Shi hshi@flatironinstitute.org}
\affiliation{Center for Computational Quantum Physics, Flatiron Institute, New York, NY 10010 USA}

\begin{abstract}
Layered perovskite ruthenium oxides exhibit a striking series of metal-insulator and magnetic-nonmagnetic  phase transitions easily tuned by temperature, pressure, epitaxy, and nonlinear drive. 
In this work, we combine results from two complementary state of the art many-body methods, Auxiliary Field Quantum Monte Carlo and Dynamical Mean Field Theory, to determine the low temperature phase diagram of Ca$_2$RuO$_4$. Both methods predict a low temperature, pressure-driven metal-insulator transition accompanied by a ferromagnetic-antiferromagnetic transition. The properties of the ferromagnetic state vary non-monotonically with pressure and are dominated by the ruthenium $d_{xy}$ orbital, while the properties of the antiferromagnetic state are dominated by the $d_{xz}$ and $d_{yz}$ orbitals.  Differences of detail in the predictions of the two methods are analyzed. This work is theoretically important as it presents the first application of the Auxiliary Field Quantum Monte Carlo method to an orbitally-degenerate system with both Mott and Hunds physics, and provides an important comparison of the Dynamical Mean Field and Auxiliary Field Quantum Monte Carlo methods. 
\end{abstract}

\maketitle

The quantum many-body problem is one of the grand challenge scientific problems of our time \cite{Tsymbal13}.  Recent work \cite{LeBlanc_PRX,Zheng_Science,Motta17} suggests that an important route towards a solution is to attack important problems via complementary methods. In this paper, we use the Auxiliary Field Quantum Monte Carlo (AFQMC) and Dynamical Mean Field Theory (DMFT) methods to study the low temperature phase diagram and physical properties  of Ca$_2$RuO$_4$.  
In the form used in this article, AFQMC is a zero temperature, finite system method that employs an imaginary time projection that samples the space of non-orthogonal Slater determinants to estimate the ground state wave function \cite{Zhang_Gubernatis_PRB_1997,Zhang_Book_2013}. In contrast, DMFT  uses a self energy  locality assumption to approximate  Green's functions at non-zero temperature \cite{Georges_RevModPhys_1996}.  The completely different natures of the approximations made and computational challenges faced by the two methods means that a comparison of results yields important insights into both the actual physics of the systems studied and the validity of the different approximations. 

The material chosen for study, Ca$_2$RuO$_4$,  is a member of a fascinating and extensively studied family of ruthenium-based compounds with chemical formulae Sr$_{n+1}$Ru$_{n}$O$_{3n+1}$ and Ca$_{n+1}$Ru$_{n}$O$_{3n+1}$. This family of materials has been of  intense interest for their remarkable properties, including unconventional  superconductivity \cite{Maeno_Lichtenberg_Nauture_1994}, variety of  magnetic phases \cite{Nakatsuji_Yoshiteru_JPSJ_1997,Cao_Guertin_PRL_1997}, nematicity,  metal-insulator transitions (MIT) \cite{Nakamura_PRB_2002,Steffens_PRB_2005}, and unusual nonequilibrium properties  \cite{Sow_Science}, all of which are believed to be due to  strong Hubbard and Hunds electron-electron interactions among the electrons in the Ru-derived $t_{2g}$ orbitals \cite{Georges_AnnRev_2013}.  

\begin{figure}[htbp]
\centering
\includegraphics[width=8cm]{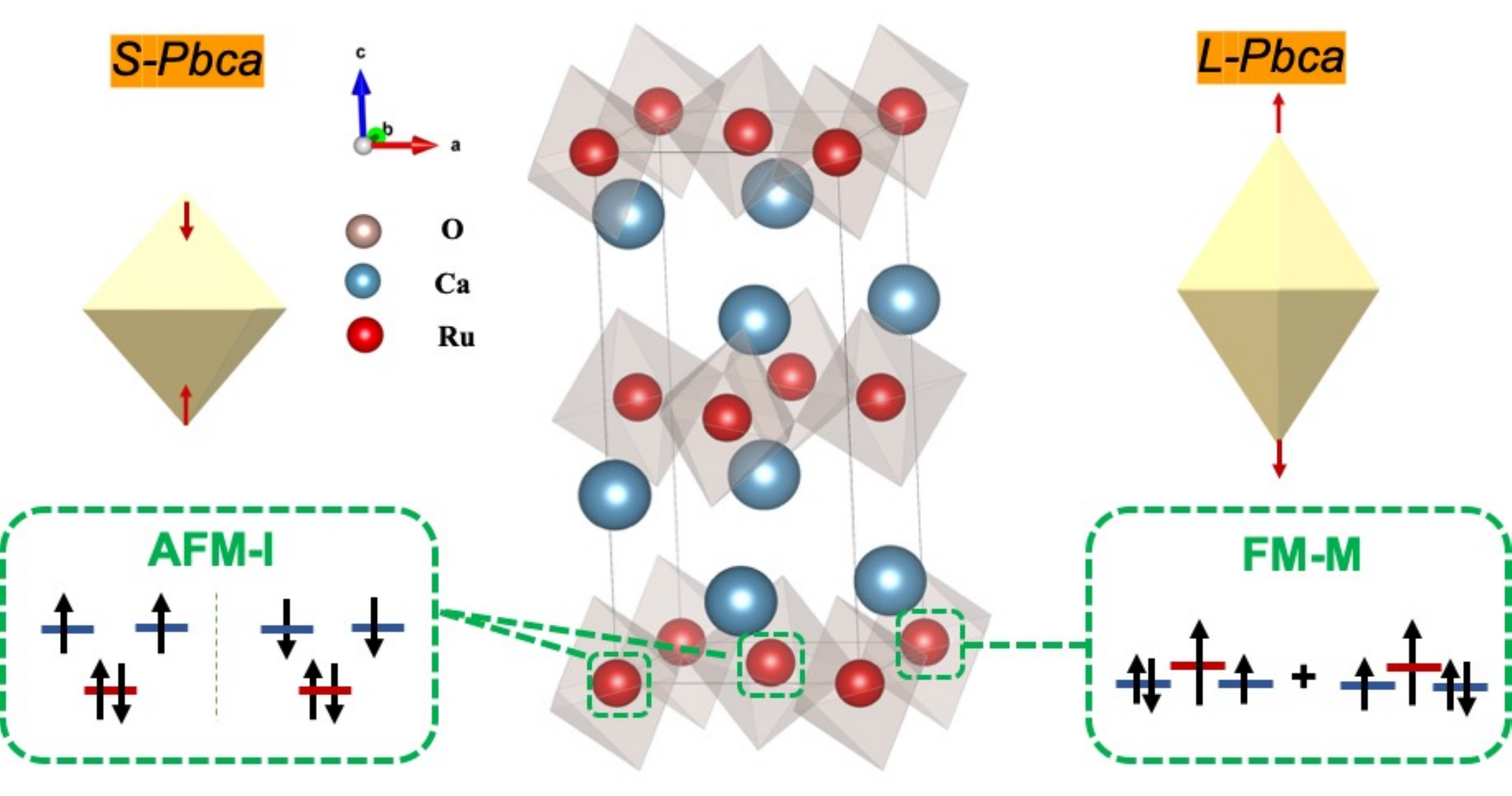}
\captionsetup{font=footnotesize, justification=raggedright, singlelinecheck=false}
\caption{Crystal structure of Ca$_{2}$RuO$_{4}$ (central portion). Calcium atoms are depicted in blue, oxygen atoms are depicted in grey, and ruthenium atoms are depicted in red. The primitive vectors $\vec{a}$, $\vec{b}$, and $\vec{c}$ are defined using the orthorhombic convention. Left side: depiction of the short-bond RuO$_6$ octahedron occurring in the low $T$, ambient pressure S-Pbca structure (top) with a depiction of the dominant Ru multiplet term in the local electronic structure of the insulating state beneath. The $d_{xy}$ orbital (red bar) is fully occupied and the $xz/yz$ orbitals are half-filled and in a high spin state. Right side: depiction of the long-bond RuO$_6$ octahedron occurring in the high $T$ or high $P$ L-Pbca structures (top) with a depiction of the dominant Ru multiplet term in the ferromagnetic state beneath. A nearly half-filled, spin-polarized $xy$ orbital (red bar) with the remaining hole is in a superposition of $xz/yz$ states.} 
\label{Ca2RuO4}
\end{figure}

Ca$_2$RuO$_4$ forms a Pbca symmetry structure derived from the $n=1$ Ruddlesden-Popper structure by rotations and tilts of the RuO$_6$ octahedra. The unit cell contains $4$ Ru ions, equivalent up to a translation and rotation of the RuO$_6$ octahedron.   At ambient pressure, the material undergoes a metal-insulator transition as temperature is decreased below $\sim$350 K and becomes antiferromagnetic below $\sim$110 K \cite{Nakatsuji_Yoshiteru_JPSJ_1997,Braden_Maeno_PRB_1998,Alexander_PRB_1999,Friedt_Maeno_PRB_2001}. Increasing pressure decreases the metal-insulator transition temperature  \cite{Nakamura_PRB_2002,Steffens_PRB_2005}, with the system remaining metallic at room temperature for pressures above $\sim$0.5 GPa \cite{Nakamura_PRB_2002,Steffens_PRB_2005} and down to very low  temperatures for $P>2$ GPa  \cite{Nakamura_JPSJ_2007}.  Low-T ferromagnetism is reported for pressures of several GPa, with $T_c$ varying from $10$-$30$ K \cite{Nakamura_JPSJ_2007}. The material may exist in two closely related forms: S-Pbca (for short) and L-Pbca, distinguished by whether the apical Ru-O bond length and c-axis lattice parameter are relatively longer (L) or shorter (S). The SPbca structure is associated with insulating and antiferromagnetic (AFM) behavior and the LPbca structure with metallic and ferromagnetic (FM) behavior. Capturing the interplay between structural and electronic properties is an important challenge for theory. 

Previous publications have studied Ca$_2$RuO$_4$ using Density Functional Theory (DFT) and its ``+U'' \cite{Anisimov_EurPhysJB_2002,Jung_PRL_2003} and plus Dynamical Mean Field Theory (DFT+DMFT) \cite{Liebsch_Ishida_PRL_2007,Gorelov_PRL_2010,Zhang_PRB_2017,Han_Millis_PRL_2018} extensions. 
However, these works primarily focused on ambient pressure phases and have each presented results from only one theoretical method. Further, $+U$ methods treat the many body physics via a Hartree-type approximation,  while DMFT, which transcends static mean field theory, makes a strong self-energy  locality assumption that may be questioned for  electronically two-dimensional materials such as the ruthenates.  For this reason, cross-comparison with another many body method, such as the AFQMC considered here, is invaluable. 

We downfold the full electronic structure of  Ca$_2$RuO$_4$ to a material-based, three-band Hamiltonian representing the correlated frontier orbitals for several different crystal structures  by first using the non-spin-polarized Generalized Gradient Approximation (GGA) \cite{Kresse_PRB_1993,Kresse_CMS_1996,Kresse_PRB_1996,Kresse_PRB_1999} to obtain an electronic band structure. We then extract frontier orbitals near the Fermi-surface states from the GGA calculations via a  maximally Localized Wannier Function construction as implemented in Wannier90 \cite{Marzari_Vanderbilt_PRB_1997,Souza_Marzari_PRB_2001}. The GGA calculations are performed using experimentally-determined atom positions obtained from ambient-pressure studies performed at room temperature and $T=400$ K  \cite{Braden_Maeno_PRB_1998,Friedt_Maeno_PRB_2001}, as well as room temperature studies performed at pressures of  $1$-$5$  GPa \cite{Steffens_PRB_2005}.  
Local Coulomb ``$U$'' and ``$J$'' interaction  terms are then added 
with $U=2.3$ eV and $J=0.35$ eV, parameters previously found to produce reliable representations of the properties and phase diagrams of perovskites including CaRuO$_3$, SrRuO$_3$, and BaRuO$_3$ \cite{Dang_Millis_PRB_2013,Dang_Millis_PRB_2015,Han_Millis_PRB_2016}, as well as  Sr$_2$RuO$_4$ \cite{Mravlje2011}
(see also Ref.~\cite{sutter_Ca2RuO4_natcomm_2017} for a direct determination of the Hund's coupling  
from photoemission measurements on Ca$_2$RuO$_4$). The resulting low energy theory is a three-band Hubbard-Kanamori Hamiltonian \cite{hao_2019_auxiliary,Kanamori_PTP_1963} 
\begin{equation}
\begin{split}
\hat{H}&= \sum_{ij \nu \nu' \sigma} t_{ij}^{\nu \nu'} \hat{c}_{i\nu \sigma}^{\dagger} \hat{c}_{j \nu'\sigma}
+ U\sum_{i\nu} \hat{n}_{i\nu\uparrow} \hat{n}_{i\nu\downarrow}\\
&+ \sum\nolimits_{\substack{i, \nu\neq \nu', \sigma\sigma'}} (U-2J-J\delta_{\sigma\sigma'}) \hat{n}_{i\nu\sigma} \hat{n}_{i\nu'\sigma'}\\
&+ J\sum\limits_{\substack{i, \nu\neq \nu'}}  (\hat{c}_{i\nu\uparrow}^{\dagger}\hat{c}_{i\nu'\downarrow}^{\dagger}\hat{c}_{i\nu\downarrow}^{}\hat{c}_{i\nu'\uparrow}^{}+\hat{c}_{i\nu\uparrow}^{\dagger} \hat{c}_{i\nu\downarrow}^{\dagger}\hat{c}_{i\nu'\downarrow}\hat{c}_{i\nu'\uparrow}).
\end{split}                     
\end{equation}
In the above, $\hat{c}_{i\nu \sigma}^{\dagger}$ creates an electron with spin $\sigma$ in Wannier state $\nu$ at lattice site $i$ and $\hat{n}_{i\nu\sigma}$ denotes the corresponding number operator. The index $\nu$ labels states derived from the Ru $t_{2g}$ symmetry $d$-orbitals (with the appropriate admixture of oxygen wave functions). The first term of the Hamiltonian describes the near Fermi surface band structure,  the second describes the intraorbital Coulomb repulsion, the third describes the interorbital Coulomb repulsion, and the last contains electron pair-hopping and exchange contributions. The \emph{ab initio} parameters $t_{ij}^{\nu \nu^\prime}$ are obtained from the Wannier analysis. The on-site $i=j$ term is a $3\times 3$ matrix parametrizing  the energy splitting between the different $t_{2g}$ symmetry $d$ orbitals. In a basis aligned with the local RuO$_6$ octahedron, $t_{i=j}$ is diagonal with two degenerate eigenvalues giving the onsite energy of the $d_{xz,yz}$ orbitals and a third eigenvalue giving the energy of the $d_{xy}$ orbital.  
The crystal-field level splitting, $\Delta = \epsilon_{yz} - \epsilon_{xy}$, is generally larger in the S-Pbca structure than in L-Pbca structures. For example, $\Delta=0.23$ eV for the ambient pressure-$295$K SPbca structure, while $\Delta=0.10$ eV for the ambient pressure-$400$K  LPbca  structure. As pressure is applied, the crystal-field splitting decreases to $\Delta=0.06$ eV for the L-$1$GPa structure and even a negative value of $\Delta=-0.02$ eV for the L-$5$GPa structure. It is important to emphasize that $\Delta$ is a ``bare'' parameter, which is small compared to the bandwidths, but whose effects may be strongly enhanced by correlations.

We treat the interactions using  the AFQMC and DMFT methods. Extending the AFQMC methodology,  which has heretofore mainly been applied to variants of the single-orbital Hubbard model, to the multiband, Hunds metal case has been an important challenge. Here, we employ the methods introduced in Ref.   \cite{hao_2019_auxiliary} to overcome this challenge.  In AFQMC, one typically studies  three-dimensional $L_x\times L_y \times L_z$ supercells. We have found that correlations along $z$ are typically very weak and therefore  set $L_z=1$ for most of the calculations. A $1\times1\times1$ unit cell contains $4$ Ru ions; the largest cell we study is  $4\times 4\times1$, containing $64$ Ru ions. 
The AFQMC method uses imaginary time propagation of a trial wave function to converge to a ground state. Our calculations use different types of trial wave functions including free-electron, as well as antiferromagnetic and ferromagnetic Hartree-Fock states. The self-consistent procedure of Ref.~\cite{PhysRevB.94.235119} is applied to find the best single-determinant trial wave function. 

We solve the three-band model employing the single shot (no charge self-consistency) DMFT \cite{Gorelov_PRL_2010,Dang_Millis_PRB_2013,Dang_Millis_PRB_2015,Han_Millis_PRB_2016,Han_Millis_PRL_2018} approximation which treats the experimental crystal structure and use the hybridization expansion variant of the continuous-time quantum Monte Carlo (CT-HYB) solver as implemented in the Toolbox for Research on Interacting Quantum Systems (TRIQS) library \cite{TRIQS_CPC_2015,TRIQS/CTHYB_CPC_2016}. Within our DMFT calculations, the single-site approximation is made and the orbital basis on each of the $4$ crystallographically inequivalent Ru sites is aligned with the local octahedral axes to minimize the sign problem.

\begin{center}
\begin{figure}[htbp]
\includegraphics[width=8cm]{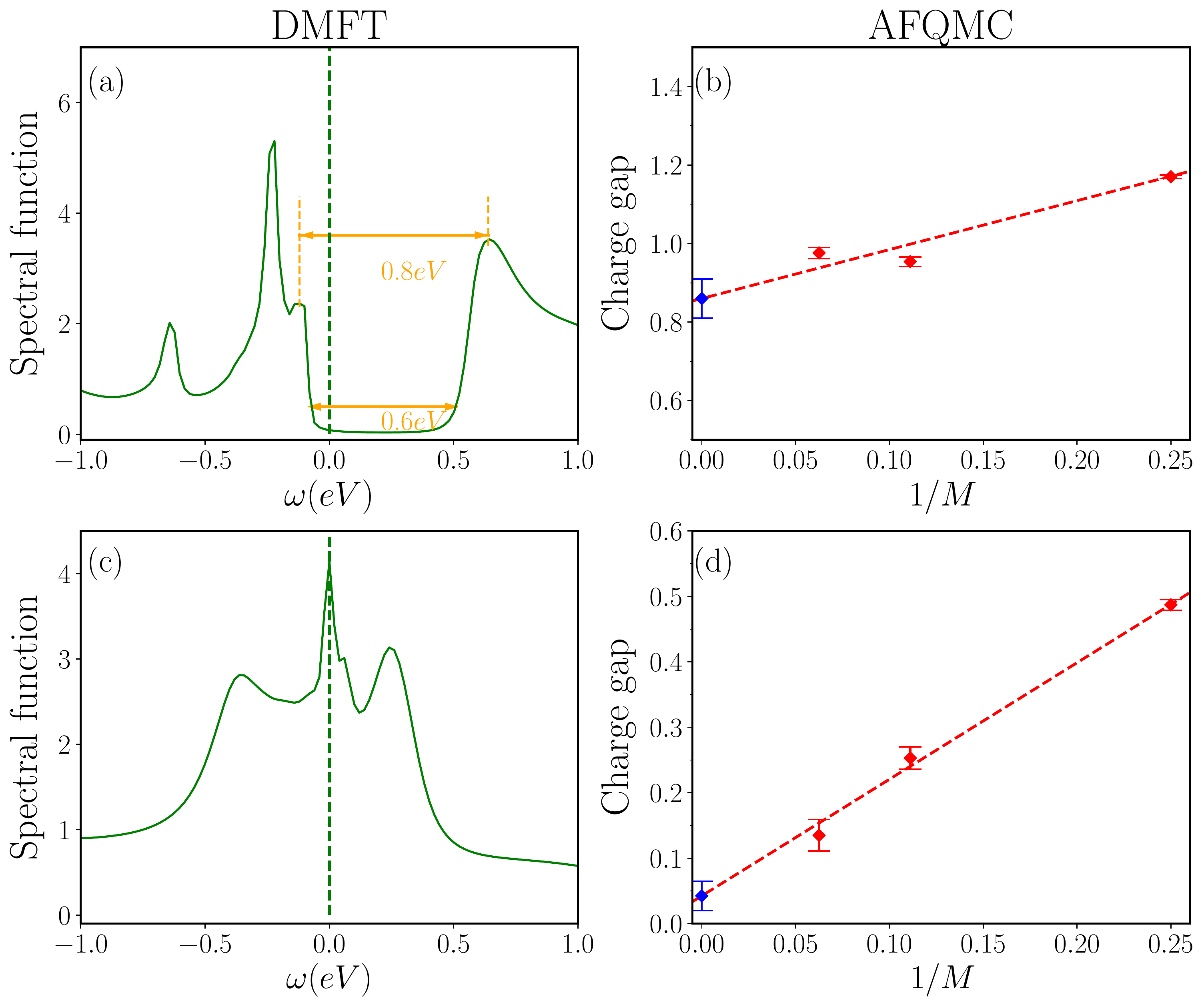}
\captionsetup{font=footnotesize, justification=raggedright, singlelinecheck=false}
\caption{Comparison of the charge gap predictions from DMFT and AFQMC for the L-Pbca and S-Pbca structures. DMFT results are calculated at $60$ K and AFQMC results are calculated at $0$ K. (a) Spectral functions per Ru obtained using DMFT simulations with analytical continuation for the S-$295$K structure. (b) Extrapolation of the charge gap to the thermodynamic limit in AFQMC calculations of the S-$295$K structure. (c) Spectral functions per Ru obtained using DMFT simulations with analytical continuation for the L-$5$GPa structure. (d) Extrapolation of the charge gap to the thermodynamic limit in AFQMC calculations of the L-$5$GPa structure. In (b) and (d), the $M$ on the $x$ axis represents the number of unit cells used in the AFQMC simulations.}
\label{Spectral_ChargeGap}
\end{figure}
\end{center}

We now present our results, beginning with the ambient-pressure, low-T S-Pbca structure. In this structure, Ca$_{2}$RuO$_{4}$ is an antiferromagnetic insulator(AFM-I) with an essentially fully occupied $d_{xy}$ orbital and half-filled $d_{xz/yz}$ orbitals~\cite{Gorelov_PRL_2010,sutter_Ca2RuO4_natcomm_2017}. 
The half-filled orbitals are in a high spin configuration and the Ru sites are antiferromagnetically ordered below a N\'eel temperature of approximately $110$ K. 
Our  calculations reproduce the observed insulating, antiferromagnetic ground state. Our AFQMC simulations are for finite-sized systems and have no spontaneous symmetry breaking, but calculations of the spin-spin correlation function reveal that the spatial extent of the correlations is at least  the size of the computational system. Our DMFT calculations were conducted at $60$ K and recover a fully polarized antiferromagnetic state and a very small imaginary self-energy.  The upper panels of Fig.~\ref{Spectral_ChargeGap} present the gap to charge excitations computed in both methods. The left panel shows the many-body density of states computed  within DMFT for the ambient pressure, $T=295$ K structure using maximum entropy analytical continuation of imaginary time quantum Monte Carlo measurements of the Green's function. The right panel shows the AFQMC charge gap computed from the difference of ground state energies having different particle numbers:
$\Delta_g=E_{N-1}+E_{N-1}-2E_{N}$. The charge gap was calculated for $2\times2\times1$, $3\times3\times1$, and $4\times4\times1$ supercells, and linear extrapolation was performed with respect to the inverse of the total number of unit cells in the computational system, revealing a $M\rightarrow\infty$  charge gap that is $\sim$0.8~eV. This is larger than the $0.6$~eV DMFT charge gap, but consistent with the energy separation between DOS maxima seen in the DMFT calculations.  Within the DMFT calculations, the physics of the insulating state is evident: from the orbitally-resolved density of states, one sees a fully occupied $xy$ band and half-occupied $xz/yz$ states with a clear  upper and lower Hubbard band structure. The near quantitatve agreement between the two calculations is strong evidence that both methods correctly represent the insulating antiferromagnetic state. The reported experimental optical gap on the antiferromagnetic phase is of the order of $0.6-0.7$~eV~\cite{Jung_PRL_2003}.

\begin{center}
\begin{figure}[htbp]
\includegraphics[width=8cm]{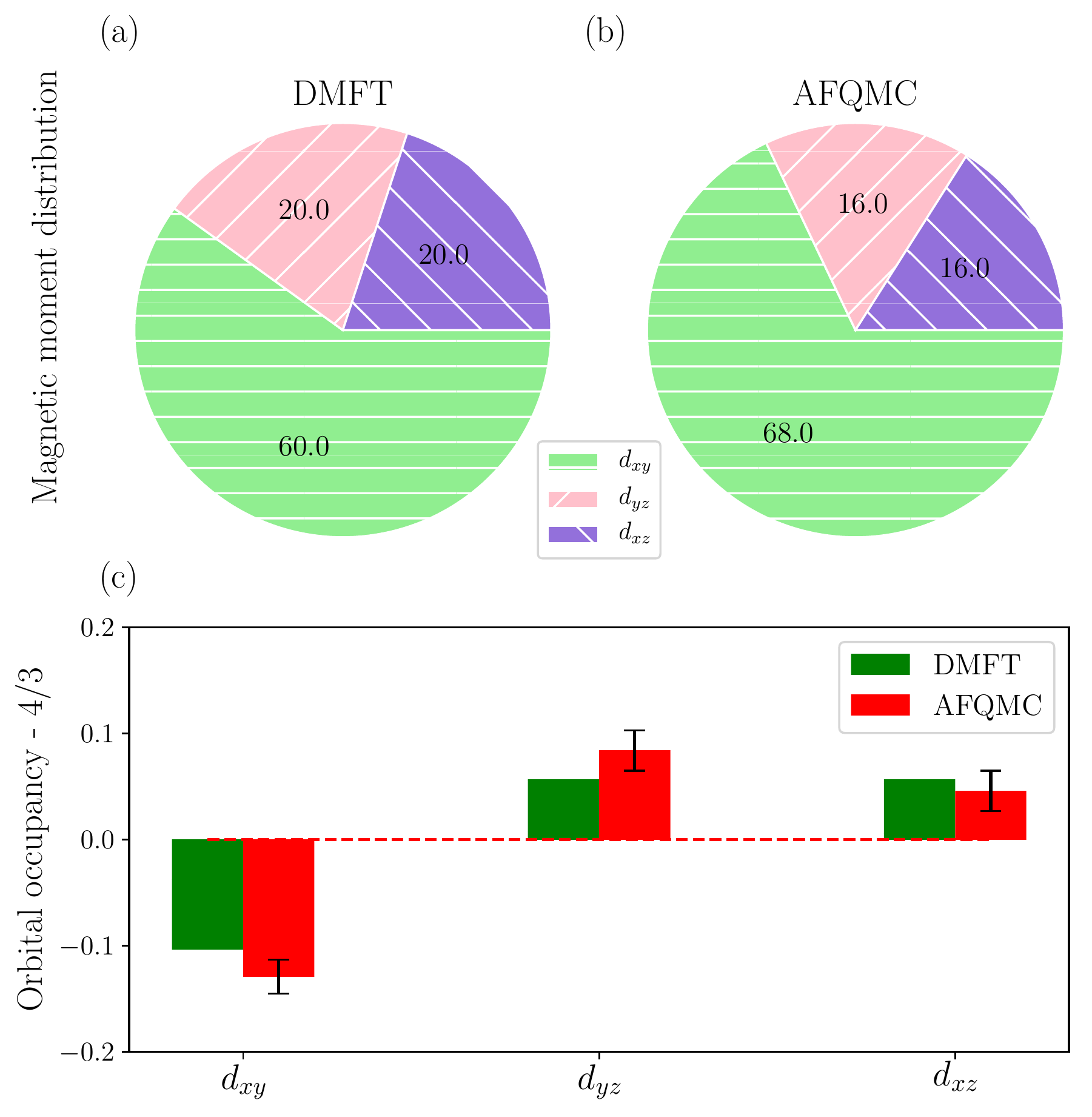}
\captionsetup{font=footnotesize, justification=raggedright, singlelinecheck=false}
\caption{Comparison of magnetic moment distributions and orbital occupancies obtained by DMFT at 60 K and AFQMC at 0 K for the experimentally acquired L-Pbca L-$5$GPa structure. (a) Magnetic moment distributions from DMFT: $60\%$ of the magnetic moment is in the $d_{xy}$ band. (b) Magnetic moment distributions from AFQMC: $68\%$ of the magnetic moment is in the $d_{xy}$ band. (c) Orbital occupancies from DMFT and AFQMC: the $y$ axis is shifted by an average density of $4/3$. A smaller $d_{xy}$ band occupancy results from a negative crystal-field splitting.}
\label{magnet_occupancy} 
\end{figure}
\end{center}

We next turn to the 5GPa L-Pbca structure, experimentally known to host a metallic state with a ferromagnetic  transition  at temperatures below $10$ K \cite{Nakamura_JPSJ_2007}. For this structure, both our DMFT and AFQMC calculations uncover a ferromagnetic metal (FM-M). The DMFT N\'eel temperature of roughly $70$ K is determined by applying a magnetic field $H$, computing the resulting magnetization, and  plotting the data according to the Arrott relation, $m^2 = c_1 H/m - c_2 (T-T_c)$. Note that, because DMFT neglects spatial fluctuations, it is expected to overestimate the ordering temperature. In order to determine the ground state magnetic order in AFQMC, we break the spin symmetry of the AFQMC trial wave function and compare the QMC energies of the different symmetry sectors. The DMFT density of states is shown in the lower left panel of Fig. \ref{Spectral_ChargeGap} and is clearly metallic. The extrapolated gap based on the AFQMC calculations is shown in the lower right panel, and is again consistent with a metallic state.  We rationalize the appearance of the metallic state by noting that the increased pressure decreases the crystal field splitting, thereby promoting the transfer of electrons from the $d_{xy}$ to the $d_{xz/yz}$ orbitals. The lower panels of Fig. \ref{magnet_occupancy} show the deviations of orbital occupancy from the equal occupancy value of $\frac{4}{3}$ for different structures; the two methods agree very well. The relatively small changes in crystal field splitting are enhanced by interaction effects, leading to almost equally occupied orbitals on the L-Pbca structure and fully occupied $d_{xy}$ orbital on the S-Pbcs structure.   


To further characterize the metallic, magnetic state, we present in the upper panels of Fig.~\ref{magnet_occupancy}  the orbital content of the magnetic moments determined from DMFT and AFQMC calculations of the L-Pbca ferromagnetic state. The two methods agree that, despite the nearly equal occupancies of the three orbitals, the dominant contribution to the ferromagnetism comes from the $d_{xy}$ orbital. The enhanced contribution of the $xy$ orbital to the magnetic moment may arise from the strong van-Hove singularity occurring near the Fermi surface in the $d_{xy}$ density of states.

\begin{figure}[htbp]
\centering
\includegraphics[width=8cm]{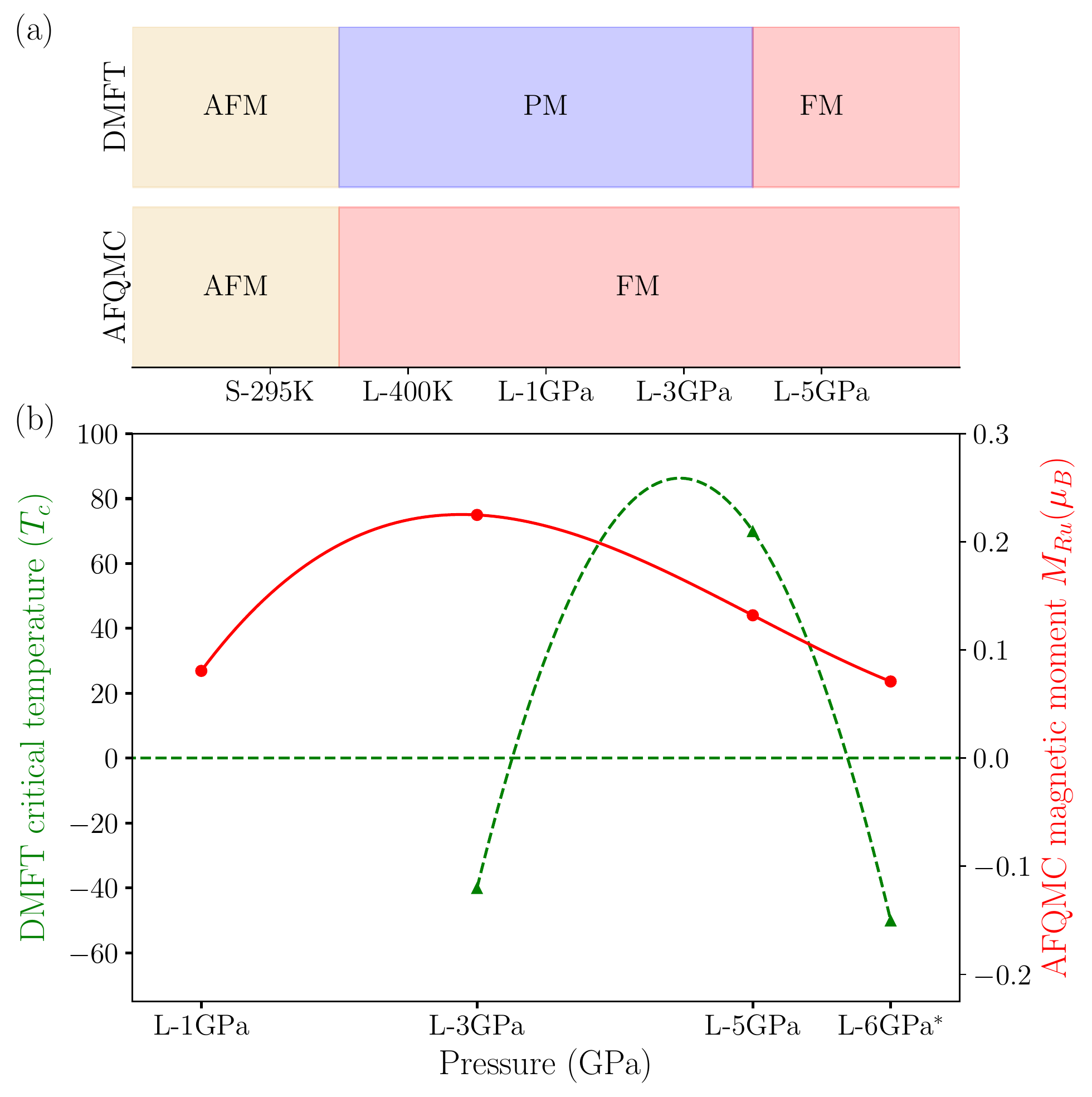}
\captionsetup{font=footnotesize, justification=raggedright, singlelinecheck=false}
\caption{(a) Phase diagram of Ca$_2$RuO$_4$ for different lattice structures. Both DMFT and AFQMC predict the S-$295$K structure to be an AFM-I 
and the L-$5$GPa structure to be an FM-M. At intermediate pressures, both DMFT and AFQMC find Ca$_2$RuO$_4$ to be a metal, but AFQMC predicts an FM-M state, while DMFT predicts a paramagnetic metal (PM-M) state (based on the range of temperatures we could access).
(b) The red line depicts the ground state magnetic moments obtained using AFQMC as a function of pressure. The green line denotes the critical temperature obtained as a function of pressure with DMFT. Pressures at which there is no FM order are depicted as having negative critical temperatures.  Note that the structure at $P=6$ GPa shown in (b) is obtained by linear extrapolation of experimentally measured atom positions at $P=1$ GPa and $P=3$ GPa.} 
\label{phase-diag}
\end{figure}

The ground state phase diagram of Ca$_2$RuO$_4$ at various pressures is depicted in Figure \ref{phase-diag}. Both AFQMC and DMFT find the S-Pbca, S-$295$K structure to be an AFM-I, consistent with experimental findings \cite{Braden_Maeno_PRB_1998,Friedt_Maeno_PRB_2001}. Both methods also find all of the L-Pbca structures studied, including the L-$400$K, L-$1$GPa, L-$3$GPa, and L-$5$GPa structures, to possess a metallic ground state and the L-$5$GPa structure to be in the FM-M state. However, AFQMC and DMFT predict different magnetic properties for many of the L-Pbca structures. 
For the L-$400$K, L-$1$GPa, and L-$3$GPa structures, AFQMC finds a FM-M state, while DMFT finds a PM-M state. Note, however, that this conclusion relies on an extrapolation from the lowest temperature that we could study ($60$~K). It will be interesting in the future to  reconsider this issue as improved DMFT solvers able to reach much lower temperatures become available. Interestingly, Ca$_2$RuO$_4$ was found to be in a mixed AFM-FM state at pressures less than $3$ GPa \cite{Lebre_Maeno_JPSJ_2007}. 
For pressures above $5$ GPa, experiments find FM order, but the critical temperature exhibits a non-monotonic behavior with pressure and peaks between 3-5 GPa in experiment \cite{Nakamura_JPSJ_2007}. We plot the critical temperature from DMFT (obtained as described from Arrott plots) as a function of pressure in Figure \ref{phase-diag}(b). A dome-shaped curve of critical temperature vs. pressure is observed,  with a peak around 5 GPa. We depict the critical temperature as being negative when DMFT does not recover a ferromagnetic state.  While the AFQMC ground state is always ferromagnetic, magnetic moments also exhibit a non-monotonic behavior, which indicates that FM order is favored at intermediate pressures. The maximum critical temperature indicated by the magnetic moment is shifted to lower pressures in AFQMC, with AFQMC's peak instead occurring around $3$ GPa. 
  
In summary, we have employed two state-of-the-art electronic structure methods,  a new implementation of AFQMC suitable for multiorbital and Hunds-coupled models,  
and DMFT, to study Ca$_2$RuO$_4$'s phase diagram and physical properties  as a function of pressure. 
Our calculations are based on a model Hamiltonian with bare electronic parameters taken from Wannier fits to density functional calculations for different experimentally reported structures. We find that the two theoretical methods yield substantially similar results for the nature of the electronic state (metal vs. insulating) and basic electronic properties including gaps and the orbital content of magnetic moments. As noted in previous work, \cite{Liebsch_Ishida_PRL_2007,Gorelov_PRL_2010,Zhang_PRB_2017,Han_Millis_PRL_2018,Sutter_Chang_NC_2017} the key to the physics is the enhancement, by interactions, 
of structurally-induced differences in the on-site splitting between the different electronic states which, although small in comparison to the overall bandwidths, drive substantial differences in the occupancies of its $t_{2g}$ orbitals.  

The  agreement between methods, occurring despite their very different approximations and sources of errors, adds confidence both to the methods and to the emerging physical picture. The most significant discrepancy between the methods is a different magnetic phase diagram for the high pressure phases of Ca$_2$RuO$_4$'s magnetic ordering between 1 and 5 GPa. The two methods agree on the qualitative features including the presence of a dome of magnetization, 
with the strongest magnetic state occurring at an intermediate pressure, but disagree on the exact range of pressures where magnetism is observed and on the pressure that maximizes the tendency toward magnetism. The difference may be due to intersite effects requiring a cluster dynamical mean field treatment in this quasi-two dimensional system, limitations in extrapolating DMFT results down to very low temperatures (calling for the development of improved solvers), or bias in AFQMC from the choice of initial wave function or the constrained path approximation. 

Our work suggests several interesting extensions. The methodology employed here is directly applicable to Ca$_{2-x}$Sr$_x$RuO$_{4}$ materials, strontium-doped versions of Ca$_2$RuO$_4$ that interpolate between the AF-I Ca$_2$RuO$_4$ studied here and metallic and superconducting  Sr$_2$RuO$_4$  \cite{Maeno_Lichtenberg_Nauture_1994}. Past works have shown that increasing $x$ is analogous to increasing temperature or pressure in this work, leading to the evolution of a metal for large values of $x$ \cite{Nakatsuji_PRL_2000, Friedt_Maeno_PRB_2001}. Nevertheless, Ca$_{2-x}$Sr$_x$RuO$_{4}$ exhibits a number of yet-to-be-explained exotic phases, including a metamagnetic phase that emerges for $0.2 < x < 0.5$ \cite{Nakatsuji_PRL_2000}. Beyond these specific applications, the methodologies employed here are ripe for application to the many 4$d$- and 5$d$-transition metal oxides whose complex interplay of spin-orbit coupling, exchange, and crystal-field effects have and continue to reveal unexpected physics. Most importantly, despite their algorithmic limitations, our new multiband AFQMC methodology and DMFT predict similar insulating and magnetic orders over wide swaths of Ca$_2$RuO$_4$'s phase diagram, only differing in their predictions of Ca$_2$RuO$_4$'s magnetic ordering for L-Pbca structures at moderate pressures. 

{\it Acknowledgements:} 
B.M.R. and H.H. acknowledge support for this work from NSF grant  DMR-1726213, DOE grant DE-SC0019441, and the Alfred P. Sloan Foundation. 
A.J.M. (partly) and Q.H. (fully) were supported by the Basic Energy Sciences Program of the US Department of Energy under grant DOE-SC-0012375. A.G. acknowledges partial support by the European Research Council (ERC-319286-QMAC). 
We are grateful to H. T. Dang for his contribution to some of the DMFT codes used in this work.
This work was conducted using computational resources and services at the Brown University Center for Computation and Visualization, XSEDE, 
the Yeti HPC cluster at Columbia University, and the Flatiron Institute. 
The Flatiron Institute is a division of the Simons Foundation.
\bibliography{references}

\begin{thebibliography}{43}%
\makeatletter
\providecommand \@ifxundefined [1]{%
 \@ifx{#1\undefined}
}%
\providecommand \@ifnum [1]{%
 \ifnum #1\expandafter \@firstoftwo
 \else \expandafter \@secondoftwo
 \fi
}%
\providecommand \@ifx [1]{%
 \ifx #1\expandafter \@firstoftwo
 \else \expandafter \@secondoftwo
 \fi
}%
\providecommand \natexlab [1]{#1}%
\providecommand \enquote  [1]{``#1''}%
\providecommand \bibnamefont  [1]{#1}%
\providecommand \bibfnamefont [1]{#1}%
\providecommand \citenamefont [1]{#1}%
\providecommand \href@noop [0]{\@secondoftwo}%
\providecommand \href [0]{\begingroup \@sanitize@url \@href}%
\providecommand \@href[1]{\@@startlink{#1}\@@href}%
\providecommand \@@href[1]{\endgroup#1\@@endlink}%
\providecommand \@sanitize@url [0]{\catcode `\\12\catcode `\$12\catcode
  `\&12\catcode `\#12\catcode `\^12\catcode `\_12\catcode `\%12\relax}%
\providecommand \@@startlink[1]{}%
\providecommand \@@endlink[0]{}%
\providecommand \url  [0]{\begingroup\@sanitize@url \@url }%
\providecommand \@url [1]{\endgroup\@href {#1}{\urlprefix }}%
\providecommand \urlprefix  [0]{URL }%
\providecommand \Eprint [0]{\href }%
\providecommand \doibase [0]{http://dx.doi.org/}%
\providecommand \selectlanguage [0]{\@gobble}%
\providecommand \bibinfo  [0]{\@secondoftwo}%
\providecommand \bibfield  [0]{\@secondoftwo}%
\providecommand \translation [1]{[#1]}%
\providecommand \BibitemOpen [0]{}%
\providecommand \bibitemStop [0]{}%
\providecommand \bibitemNoStop [0]{.\EOS\space}%
\providecommand \EOS [0]{\spacefactor3000\relax}%
\providecommand \BibitemShut  [1]{\csname bibitem#1\endcsname}%
\let\auto@bib@innerbib\@empty
\bibitem [{\citenamefont {Tsymbal}\ and\ \citenamefont
  {Dowben}(2013)}]{Tsymbal13}%
  \BibitemOpen
  \bibfield  {author} {\bibinfo {author} {\bibfnamefont {E.}~\bibnamefont
  {Tsymbal}}\ and\ \bibinfo {author} {\bibfnamefont {P.}~\bibnamefont
  {Dowben}},\ }\href {\doibase 10.3389/fphy.2013.00032} {\bibfield  {journal}
  {\bibinfo  {journal} {Front. Phys.}\ }\textbf {\bibinfo {volume} {1}},\
  \bibinfo {pages} {32} (\bibinfo {year} {2013})}\BibitemShut {NoStop}%
\bibitem [{\citenamefont {LeBlanc}\ \emph {et~al.}(2015)\citenamefont
  {LeBlanc}, \citenamefont {Antipov}, \citenamefont {Becca}, \citenamefont
  {Bulik}, \citenamefont {Chan}, \citenamefont {Chung}, \citenamefont {Deng},
  \citenamefont {Ferrero}, \citenamefont {Henderson}, \citenamefont
  {Jim\'enez-Hoyos}, \citenamefont {Kozik}, \citenamefont {Liu}, \citenamefont
  {Millis}, \citenamefont {Prokof'ev}, \citenamefont {Qin}, \citenamefont
  {Scuseria}, \citenamefont {Shi}, \citenamefont {Svistunov}, \citenamefont
  {Tocchio}, \citenamefont {Tupitsyn}, \citenamefont {White}, \citenamefont
  {Zhang}, \citenamefont {Zheng}, \citenamefont {Zhu},\ and\ \citenamefont
  {Gull}}]{LeBlanc_PRX}%
  \BibitemOpen
  \bibfield  {author} {\bibinfo {author} {\bibfnamefont {J.~P.~F.}\
  \bibnamefont {LeBlanc}}, \bibinfo {author} {\bibfnamefont {A.~E.}\
  \bibnamefont {Antipov}}, \bibinfo {author} {\bibfnamefont {F.}~\bibnamefont
  {Becca}}, \bibinfo {author} {\bibfnamefont {I.~W.}\ \bibnamefont {Bulik}},
  \bibinfo {author} {\bibfnamefont {G.~K.-L.}\ \bibnamefont {Chan}}, \bibinfo
  {author} {\bibfnamefont {C.-M.}\ \bibnamefont {Chung}}, \bibinfo {author}
  {\bibfnamefont {Y.}~\bibnamefont {Deng}}, \bibinfo {author} {\bibfnamefont
  {M.}~\bibnamefont {Ferrero}}, \bibinfo {author} {\bibfnamefont {T.~M.}\
  \bibnamefont {Henderson}}, \bibinfo {author} {\bibfnamefont {C.~A.}\
  \bibnamefont {Jim\'enez-Hoyos}}, \bibinfo {author} {\bibfnamefont
  {E.}~\bibnamefont {Kozik}}, \bibinfo {author} {\bibfnamefont {X.-W.}\
  \bibnamefont {Liu}}, \bibinfo {author} {\bibfnamefont {A.~J.}\ \bibnamefont
  {Millis}}, \bibinfo {author} {\bibfnamefont {N.~V.}\ \bibnamefont
  {Prokof'ev}}, \bibinfo {author} {\bibfnamefont {M.}~\bibnamefont {Qin}},
  \bibinfo {author} {\bibfnamefont {G.~E.}\ \bibnamefont {Scuseria}}, \bibinfo
  {author} {\bibfnamefont {H.}~\bibnamefont {Shi}}, \bibinfo {author}
  {\bibfnamefont {B.~V.}\ \bibnamefont {Svistunov}}, \bibinfo {author}
  {\bibfnamefont {L.~F.}\ \bibnamefont {Tocchio}}, \bibinfo {author}
  {\bibfnamefont {I.~S.}\ \bibnamefont {Tupitsyn}}, \bibinfo {author}
  {\bibfnamefont {S.~R.}\ \bibnamefont {White}}, \bibinfo {author}
  {\bibfnamefont {S.}~\bibnamefont {Zhang}}, \bibinfo {author} {\bibfnamefont
  {B.-X.}\ \bibnamefont {Zheng}}, \bibinfo {author} {\bibfnamefont
  {Z.}~\bibnamefont {Zhu}}, \ and\ \bibinfo {author} {\bibfnamefont
  {E.}~\bibnamefont {Gull}} (\bibinfo {collaboration} {Simons Collaboration on
  the Many-Electron Problem}),\ }\href {\doibase 10.1103/PhysRevX.5.041041}
  {\bibfield  {journal} {\bibinfo  {journal} {Phys. Rev. X}\ }\textbf {\bibinfo
  {volume} {5}},\ \bibinfo {pages} {041041} (\bibinfo {year}
  {2015})}\BibitemShut {NoStop}%
\bibitem [{\citenamefont {Zheng}\ \emph {et~al.}(2017)\citenamefont {Zheng},
  \citenamefont {Chung}, \citenamefont {Corboz}, \citenamefont {Ehlers},
  \citenamefont {Qin}, \citenamefont {Noack}, \citenamefont {Shi},
  \citenamefont {White}, \citenamefont {Zhang},\ and\ \citenamefont
  {Chan}}]{Zheng_Science}%
  \BibitemOpen
  \bibfield  {author} {\bibinfo {author} {\bibfnamefont {B.-X.}\ \bibnamefont
  {Zheng}}, \bibinfo {author} {\bibfnamefont {C.-M.}\ \bibnamefont {Chung}},
  \bibinfo {author} {\bibfnamefont {P.}~\bibnamefont {Corboz}}, \bibinfo
  {author} {\bibfnamefont {G.}~\bibnamefont {Ehlers}}, \bibinfo {author}
  {\bibfnamefont {M.-P.}\ \bibnamefont {Qin}}, \bibinfo {author} {\bibfnamefont
  {R.~M.}\ \bibnamefont {Noack}}, \bibinfo {author} {\bibfnamefont
  {H.}~\bibnamefont {Shi}}, \bibinfo {author} {\bibfnamefont {S.~R.}\
  \bibnamefont {White}}, \bibinfo {author} {\bibfnamefont {S.}~\bibnamefont
  {Zhang}}, \ and\ \bibinfo {author} {\bibfnamefont {G.~K.-L.}\ \bibnamefont
  {Chan}},\ }\href {\doibase 10.1126/science.aam7127} {\bibfield  {journal}
  {\bibinfo  {journal} {Science}\ }\textbf {\bibinfo {volume} {358}},\ \bibinfo
  {pages} {1155} (\bibinfo {year} {2017})}\BibitemShut {NoStop}%
\bibitem [{\citenamefont {Motta}\ \emph {et~al.}(2017)\citenamefont {Motta},
  \citenamefont {Ceperley}, \citenamefont {Chan}, \citenamefont {Gomez},
  \citenamefont {Gull}, \citenamefont {Guo}, \citenamefont {Jim\'enez-Hoyos},
  \citenamefont {Lan}, \citenamefont {Li}, \citenamefont {Ma}, \citenamefont
  {Millis}, \citenamefont {Prokof'ev}, \citenamefont {Ray}, \citenamefont
  {Scuseria}, \citenamefont {Sorella}, \citenamefont {Stoudenmire},
  \citenamefont {Sun}, \citenamefont {Tupitsyn}, \citenamefont {White},
  \citenamefont {Zgid},\ and\ \citenamefont {Zhang}}]{Motta17}%
  \BibitemOpen
  \bibfield  {author} {\bibinfo {author} {\bibfnamefont {M.}~\bibnamefont
  {Motta}}, \bibinfo {author} {\bibfnamefont {D.~M.}\ \bibnamefont {Ceperley}},
  \bibinfo {author} {\bibfnamefont {G.~K.-L.}\ \bibnamefont {Chan}}, \bibinfo
  {author} {\bibfnamefont {J.~A.}\ \bibnamefont {Gomez}}, \bibinfo {author}
  {\bibfnamefont {E.}~\bibnamefont {Gull}}, \bibinfo {author} {\bibfnamefont
  {S.}~\bibnamefont {Guo}}, \bibinfo {author} {\bibfnamefont {C.~A.}\
  \bibnamefont {Jim\'enez-Hoyos}}, \bibinfo {author} {\bibfnamefont {T.~N.}\
  \bibnamefont {Lan}}, \bibinfo {author} {\bibfnamefont {J.}~\bibnamefont
  {Li}}, \bibinfo {author} {\bibfnamefont {F.}~\bibnamefont {Ma}}, \bibinfo
  {author} {\bibfnamefont {A.~J.}\ \bibnamefont {Millis}}, \bibinfo {author}
  {\bibfnamefont {N.~V.}\ \bibnamefont {Prokof'ev}}, \bibinfo {author}
  {\bibfnamefont {U.}~\bibnamefont {Ray}}, \bibinfo {author} {\bibfnamefont
  {G.~E.}\ \bibnamefont {Scuseria}}, \bibinfo {author} {\bibfnamefont
  {S.}~\bibnamefont {Sorella}}, \bibinfo {author} {\bibfnamefont {E.~M.}\
  \bibnamefont {Stoudenmire}}, \bibinfo {author} {\bibfnamefont
  {Q.}~\bibnamefont {Sun}}, \bibinfo {author} {\bibfnamefont {I.~S.}\
  \bibnamefont {Tupitsyn}}, \bibinfo {author} {\bibfnamefont {S.~R.}\
  \bibnamefont {White}}, \bibinfo {author} {\bibfnamefont {D.}~\bibnamefont
  {Zgid}}, \ and\ \bibinfo {author} {\bibfnamefont {S.}~\bibnamefont {Zhang}}
  (\bibinfo {collaboration} {Simons Collaboration on the Many-Electron
  Problem}),\ }\href {\doibase 10.1103/PhysRevX.7.031059} {\bibfield  {journal}
  {\bibinfo  {journal} {Phys. Rev. X}\ }\textbf {\bibinfo {volume} {7}},\
  \bibinfo {pages} {031059} (\bibinfo {year} {2017})}\BibitemShut {NoStop}%
\bibitem [{\citenamefont {Zhang}\ \emph {et~al.}(1997)\citenamefont {Zhang},
  \citenamefont {Carlson},\ and\ \citenamefont
  {Gubernatis}}]{Zhang_Gubernatis_PRB_1997}%
  \BibitemOpen
  \bibfield  {author} {\bibinfo {author} {\bibfnamefont {S.}~\bibnamefont
  {Zhang}}, \bibinfo {author} {\bibfnamefont {J.}~\bibnamefont {Carlson}}, \
  and\ \bibinfo {author} {\bibfnamefont {J.~E.}\ \bibnamefont {Gubernatis}},\
  }\href {\doibase 10.1103/PhysRevB.55.7464} {\bibfield  {journal} {\bibinfo
  {journal} {Phys. Rev. B}\ }\textbf {\bibinfo {volume} {55}},\ \bibinfo
  {pages} {7464} (\bibinfo {year} {1997})}\BibitemShut {NoStop}%
\bibitem [{\citenamefont {Zhang}(2013)}]{Zhang_Book_2013}%
  \BibitemOpen
  \bibfield  {author} {\bibinfo {author} {\bibfnamefont {S.}~\bibnamefont
  {Zhang}},\ }\enquote {\bibinfo {title} {Auxiliary-field quantum monte carlo
  for correlated electron systems},}\ \ (\bibinfo  {publisher} {Verlag},\
  \bibinfo {year} {2013})\ Chap.\ \bibinfo {chapter} {Emergent Phenomena in
  Correlated Matter: Modeling and Simulation}\BibitemShut {NoStop}%
\bibitem [{\citenamefont {Georges}\ \emph {et~al.}(1996)\citenamefont
  {Georges}, \citenamefont {Kotliar}, \citenamefont {Krauth},\ and\
  \citenamefont {Rozenberg}}]{Georges_RevModPhys_1996}%
  \BibitemOpen
  \bibfield  {author} {\bibinfo {author} {\bibfnamefont {A.}~\bibnamefont
  {Georges}}, \bibinfo {author} {\bibfnamefont {G.}~\bibnamefont {Kotliar}},
  \bibinfo {author} {\bibfnamefont {W.}~\bibnamefont {Krauth}}, \ and\ \bibinfo
  {author} {\bibfnamefont {M.~J.}\ \bibnamefont {Rozenberg}},\ }\href {\doibase
  10.1103/RevModPhys.68.13} {\bibfield  {journal} {\bibinfo  {journal} {Rev.
  Mod. Phys.}\ }\textbf {\bibinfo {volume} {68}},\ \bibinfo {pages} {13}
  (\bibinfo {year} {1996})}\BibitemShut {NoStop}%
\bibitem [{\citenamefont {Maeno}\ \emph {et~al.}(1994)\citenamefont {Maeno},
  \citenamefont {Hashimoto}, \citenamefont {Yoshida}, \citenamefont
  {Nishizaki}, \citenamefont {Fujita}, \citenamefont {Bednorz},\ and\
  \citenamefont {Lichtenberg}}]{Maeno_Lichtenberg_Nauture_1994}%
  \BibitemOpen
  \bibfield  {author} {\bibinfo {author} {\bibfnamefont {Y.}~\bibnamefont
  {Maeno}}, \bibinfo {author} {\bibfnamefont {H.}~\bibnamefont {Hashimoto}},
  \bibinfo {author} {\bibfnamefont {K.}~\bibnamefont {Yoshida}}, \bibinfo
  {author} {\bibfnamefont {S.}~\bibnamefont {Nishizaki}}, \bibinfo {author}
  {\bibfnamefont {T.}~\bibnamefont {Fujita}}, \bibinfo {author} {\bibfnamefont
  {J.}~\bibnamefont {Bednorz}}, \ and\ \bibinfo {author} {\bibfnamefont
  {F.}~\bibnamefont {Lichtenberg}},\ }\href {https://doi.org/10.1038/372532a0}
  {\bibfield  {journal} {\bibinfo  {journal} {Nature}\ }\textbf {\bibinfo
  {volume} {372}},\ \bibinfo {pages} {532} (\bibinfo {year}
  {1994})}\BibitemShut {NoStop}%
\bibitem [{\citenamefont {Nakatsuji}\ \emph {et~al.}(1997)\citenamefont
  {Nakatsuji}, \citenamefont {Ikeda},\ and\ \citenamefont
  {Maeno}}]{Nakatsuji_Yoshiteru_JPSJ_1997}%
  \BibitemOpen
  \bibfield  {author} {\bibinfo {author} {\bibfnamefont {S.}~\bibnamefont
  {Nakatsuji}}, \bibinfo {author} {\bibfnamefont {S.-i.}\ \bibnamefont
  {Ikeda}}, \ and\ \bibinfo {author} {\bibfnamefont {Y.}~\bibnamefont
  {Maeno}},\ }\href {https://doi.org/10.1143/JPSJ.66.1868} {\bibfield
  {journal} {\bibinfo  {journal} {J. Phys. Soc. Jpn.}\ }\textbf {\bibinfo
  {volume} {66}},\ \bibinfo {pages} {1868} (\bibinfo {year}
  {1997})}\BibitemShut {NoStop}%
\bibitem [{\citenamefont {Cao}\ \emph {et~al.}(1997)\citenamefont {Cao},
  \citenamefont {McCall}, \citenamefont {Crow},\ and\ \citenamefont
  {Guertin}}]{Cao_Guertin_PRL_1997}%
  \BibitemOpen
  \bibfield  {author} {\bibinfo {author} {\bibfnamefont {G.}~\bibnamefont
  {Cao}}, \bibinfo {author} {\bibfnamefont {S.}~\bibnamefont {McCall}},
  \bibinfo {author} {\bibfnamefont {J.}~\bibnamefont {Crow}}, \ and\ \bibinfo
  {author} {\bibfnamefont {R.}~\bibnamefont {Guertin}},\ }\href {\doibase
  10.1103/PhysRevLett.78.1751} {\bibfield  {journal} {\bibinfo  {journal}
  {Phys. Rev. Lett.}\ }\textbf {\bibinfo {volume} {78}},\ \bibinfo {pages}
  {1751} (\bibinfo {year} {1997})}\BibitemShut {NoStop}%
\bibitem [{\citenamefont {Nakamura}\ \emph {et~al.}(2002)\citenamefont
  {Nakamura}, \citenamefont {Goko}, \citenamefont {Ito}, \citenamefont
  {Fujita}, \citenamefont {Nakatsuji}, \citenamefont {Fukazawa}, \citenamefont
  {Maeno}, \citenamefont {Alireza}, \citenamefont {Forsythe},\ and\
  \citenamefont {Julian}}]{Nakamura_PRB_2002}%
  \BibitemOpen
  \bibfield  {author} {\bibinfo {author} {\bibfnamefont {F.}~\bibnamefont
  {Nakamura}}, \bibinfo {author} {\bibfnamefont {T.}~\bibnamefont {Goko}},
  \bibinfo {author} {\bibfnamefont {M.}~\bibnamefont {Ito}}, \bibinfo {author}
  {\bibfnamefont {T.}~\bibnamefont {Fujita}}, \bibinfo {author} {\bibfnamefont
  {S.}~\bibnamefont {Nakatsuji}}, \bibinfo {author} {\bibfnamefont
  {H.}~\bibnamefont {Fukazawa}}, \bibinfo {author} {\bibfnamefont
  {Y.}~\bibnamefont {Maeno}}, \bibinfo {author} {\bibfnamefont
  {P.}~\bibnamefont {Alireza}}, \bibinfo {author} {\bibfnamefont
  {D.}~\bibnamefont {Forsythe}}, \ and\ \bibinfo {author} {\bibfnamefont
  {S.~R.}\ \bibnamefont {Julian}},\ }\href {\doibase
  10.1103/PhysRevB.65.220402} {\bibfield  {journal} {\bibinfo  {journal} {Phys.
  Rev. B}\ }\textbf {\bibinfo {volume} {65}},\ \bibinfo {pages} {220402}
  (\bibinfo {year} {2002})}\BibitemShut {NoStop}%
\bibitem [{\citenamefont {Steffens}\ \emph {et~al.}(2005)\citenamefont
  {Steffens}, \citenamefont {Friedt}, \citenamefont {Alireza}, \citenamefont
  {Marshall}, \citenamefont {Schmidt}, \citenamefont {Nakamura}, \citenamefont
  {Nakatsuji}, \citenamefont {Maeno}, \citenamefont {Lengsdorf}, \citenamefont
  {Abd-Elmeguid},\ and\ \citenamefont {Braden}}]{Steffens_PRB_2005}%
  \BibitemOpen
  \bibfield  {author} {\bibinfo {author} {\bibfnamefont {P.}~\bibnamefont
  {Steffens}}, \bibinfo {author} {\bibfnamefont {O.}~\bibnamefont {Friedt}},
  \bibinfo {author} {\bibfnamefont {P.}~\bibnamefont {Alireza}}, \bibinfo
  {author} {\bibfnamefont {W.~G.}\ \bibnamefont {Marshall}}, \bibinfo {author}
  {\bibfnamefont {W.}~\bibnamefont {Schmidt}}, \bibinfo {author} {\bibfnamefont
  {F.}~\bibnamefont {Nakamura}}, \bibinfo {author} {\bibfnamefont
  {S.}~\bibnamefont {Nakatsuji}}, \bibinfo {author} {\bibfnamefont
  {Y.}~\bibnamefont {Maeno}}, \bibinfo {author} {\bibfnamefont
  {R.}~\bibnamefont {Lengsdorf}}, \bibinfo {author} {\bibfnamefont {M.~M.}\
  \bibnamefont {Abd-Elmeguid}}, \ and\ \bibinfo {author} {\bibfnamefont
  {M.}~\bibnamefont {Braden}},\ }\href {\doibase 10.1103/PhysRevB.72.094104}
  {\bibfield  {journal} {\bibinfo  {journal} {Phys. Rev. B}\ }\textbf {\bibinfo
  {volume} {72}},\ \bibinfo {pages} {094104} (\bibinfo {year}
  {2005})}\BibitemShut {NoStop}%
\bibitem [{\citenamefont {Sow}\ \emph {et~al.}(2017)\citenamefont {Sow},
  \citenamefont {Yonezawa}, \citenamefont {Kitamura}, \citenamefont {Oka},
  \citenamefont {Kuroki}, \citenamefont {Nakamura},\ and\ \citenamefont
  {Maeno}}]{Sow_Science}%
  \BibitemOpen
  \bibfield  {author} {\bibinfo {author} {\bibfnamefont {C.}~\bibnamefont
  {Sow}}, \bibinfo {author} {\bibfnamefont {S.}~\bibnamefont {Yonezawa}},
  \bibinfo {author} {\bibfnamefont {S.}~\bibnamefont {Kitamura}}, \bibinfo
  {author} {\bibfnamefont {T.}~\bibnamefont {Oka}}, \bibinfo {author}
  {\bibfnamefont {K.}~\bibnamefont {Kuroki}}, \bibinfo {author} {\bibfnamefont
  {F.}~\bibnamefont {Nakamura}}, \ and\ \bibinfo {author} {\bibfnamefont
  {Y.}~\bibnamefont {Maeno}},\ }\href {\doibase 10.1126/science.aah4297}
  {\bibfield  {journal} {\bibinfo  {journal} {Science}\ }\textbf {\bibinfo
  {volume} {358}},\ \bibinfo {pages} {1084} (\bibinfo {year}
  {2017})}\BibitemShut {NoStop}%
\bibitem [{\citenamefont {Georges}\ \emph {et~al.}(2013)\citenamefont
  {Georges}, \citenamefont {Medici},\ and\ \citenamefont
  {Mravlje}}]{Georges_AnnRev_2013}%
  \BibitemOpen
  \bibfield  {author} {\bibinfo {author} {\bibfnamefont {A.}~\bibnamefont
  {Georges}}, \bibinfo {author} {\bibfnamefont {L.~d.}\ \bibnamefont {Medici}},
  \ and\ \bibinfo {author} {\bibfnamefont {J.}~\bibnamefont {Mravlje}},\
  }\href@noop {} {\bibfield  {journal} {\bibinfo  {journal} {Annu. Rev.
  Condens. Matter Phys.}\ }\textbf {\bibinfo {volume} {4}},\ \bibinfo {pages}
  {137} (\bibinfo {year} {2013})}\BibitemShut {NoStop}%
\bibitem [{\citenamefont {Braden}\ \emph {et~al.}(1998)\citenamefont {Braden},
  \citenamefont {Andr\'e}, \citenamefont {Nakatsuji},\ and\ \citenamefont
  {Maeno}}]{Braden_Maeno_PRB_1998}%
  \BibitemOpen
  \bibfield  {author} {\bibinfo {author} {\bibfnamefont {M.}~\bibnamefont
  {Braden}}, \bibinfo {author} {\bibfnamefont {G.}~\bibnamefont {Andr\'e}},
  \bibinfo {author} {\bibfnamefont {S.}~\bibnamefont {Nakatsuji}}, \ and\
  \bibinfo {author} {\bibfnamefont {Y.}~\bibnamefont {Maeno}},\ }\href
  {\doibase 10.1103/PhysRevB.58.847} {\bibfield  {journal} {\bibinfo  {journal}
  {Phys. Rev. B}\ }\textbf {\bibinfo {volume} {58}},\ \bibinfo {pages} {847}
  (\bibinfo {year} {1998})}\BibitemShut {NoStop}%
\bibitem [{\citenamefont {Alexander}\ \emph {et~al.}(1999)\citenamefont
  {Alexander}, \citenamefont {Cao}, \citenamefont {Dobrosavljevic},
  \citenamefont {McCall}, \citenamefont {Crow}, \citenamefont {Lochner},\ and\
  \citenamefont {Guertin}}]{Alexander_PRB_1999}%
  \BibitemOpen
  \bibfield  {author} {\bibinfo {author} {\bibfnamefont {C.~S.}\ \bibnamefont
  {Alexander}}, \bibinfo {author} {\bibfnamefont {G.}~\bibnamefont {Cao}},
  \bibinfo {author} {\bibfnamefont {V.}~\bibnamefont {Dobrosavljevic}},
  \bibinfo {author} {\bibfnamefont {S.}~\bibnamefont {McCall}}, \bibinfo
  {author} {\bibfnamefont {J.~E.}\ \bibnamefont {Crow}}, \bibinfo {author}
  {\bibfnamefont {E.}~\bibnamefont {Lochner}}, \ and\ \bibinfo {author}
  {\bibfnamefont {R.~P.}\ \bibnamefont {Guertin}},\ }\href {\doibase
  10.1103/PhysRevB.60.R8422} {\bibfield  {journal} {\bibinfo  {journal} {Phys.
  Rev. B}\ }\textbf {\bibinfo {volume} {60}},\ \bibinfo {pages} {R8422}
  (\bibinfo {year} {1999})}\BibitemShut {NoStop}%
\bibitem [{\citenamefont {Friedt}\ \emph {et~al.}(2001)\citenamefont {Friedt},
  \citenamefont {Braden}, \citenamefont {Andr\'e}, \citenamefont {Adelmann},
  \citenamefont {Nakatsuji},\ and\ \citenamefont
  {Maeno}}]{Friedt_Maeno_PRB_2001}%
  \BibitemOpen
  \bibfield  {author} {\bibinfo {author} {\bibfnamefont {O.}~\bibnamefont
  {Friedt}}, \bibinfo {author} {\bibfnamefont {M.}~\bibnamefont {Braden}},
  \bibinfo {author} {\bibfnamefont {G.}~\bibnamefont {Andr\'e}}, \bibinfo
  {author} {\bibfnamefont {P.}~\bibnamefont {Adelmann}}, \bibinfo {author}
  {\bibfnamefont {S.}~\bibnamefont {Nakatsuji}}, \ and\ \bibinfo {author}
  {\bibfnamefont {Y.}~\bibnamefont {Maeno}},\ }\href {\doibase
  10.1103/PhysRevB.63.174432} {\bibfield  {journal} {\bibinfo  {journal} {Phys.
  Rev. B}\ }\textbf {\bibinfo {volume} {63}},\ \bibinfo {pages} {174432}
  (\bibinfo {year} {2001})}\BibitemShut {NoStop}%
\bibitem [{\citenamefont {Nakamura}(2007)}]{Nakamura_JPSJ_2007}%
  \BibitemOpen
  \bibfield  {author} {\bibinfo {author} {\bibfnamefont {F.}~\bibnamefont
  {Nakamura}},\ }\href {https://doi.org/10.1143/JPSJS.76SA.96} {\bibfield
  {journal} {\bibinfo  {journal} {J. Phys. Soc. Jpn.}\ }\textbf {\bibinfo
  {volume} {76}},\ \bibinfo {pages} {96} (\bibinfo {year} {2007})}\BibitemShut
  {NoStop}%
\bibitem [{\citenamefont {{Anisimov, V. I.}}\ \emph {et~al.}(2002)\citenamefont
  {{Anisimov, V. I.}}, \citenamefont {{Nekrasov, I. A.}}, \citenamefont
  {{Kondakov, D. E.}}, \citenamefont {{Rice, T. M.}},\ and\ \citenamefont
  {{Sigrist, M.}}}]{Anisimov_EurPhysJB_2002}%
  \BibitemOpen
  \bibfield  {author} {\bibinfo {author} {\bibnamefont {{Anisimov, V. I.}}},
  \bibinfo {author} {\bibnamefont {{Nekrasov, I. A.}}}, \bibinfo {author}
  {\bibnamefont {{Kondakov, D. E.}}}, \bibinfo {author} {\bibnamefont {{Rice,
  T. M.}}}, \ and\ \bibinfo {author} {\bibnamefont {{Sigrist, M.}}},\ }\href
  {https://doi.org/10.1140/epjb/e20020021} {\bibfield  {journal} {\bibinfo
  {journal} {Eur. Phys. J. B}\ }\textbf {\bibinfo {volume} {25}},\ \bibinfo
  {pages} {191} (\bibinfo {year} {2002})}\BibitemShut {NoStop}%
\bibitem [{\citenamefont {Jung}\ \emph {et~al.}(2003)\citenamefont {Jung},
  \citenamefont {Fang}, \citenamefont {He}, \citenamefont {Kaneko},
  \citenamefont {Okimoto},\ and\ \citenamefont {Tokura}}]{Jung_PRL_2003}%
  \BibitemOpen
  \bibfield  {author} {\bibinfo {author} {\bibfnamefont {J.~H.}\ \bibnamefont
  {Jung}}, \bibinfo {author} {\bibfnamefont {Z.}~\bibnamefont {Fang}}, \bibinfo
  {author} {\bibfnamefont {J.~P.}\ \bibnamefont {He}}, \bibinfo {author}
  {\bibfnamefont {Y.}~\bibnamefont {Kaneko}}, \bibinfo {author} {\bibfnamefont
  {Y.}~\bibnamefont {Okimoto}}, \ and\ \bibinfo {author} {\bibfnamefont
  {Y.}~\bibnamefont {Tokura}},\ }\href {\doibase 10.1103/PhysRevLett.91.056403}
  {\bibfield  {journal} {\bibinfo  {journal} {Phys. Rev. Lett.}\ }\textbf
  {\bibinfo {volume} {91}},\ \bibinfo {pages} {056403} (\bibinfo {year}
  {2003})}\BibitemShut {NoStop}%
\bibitem [{\citenamefont {Liebsch}\ and\ \citenamefont
  {Ishida}(2007)}]{Liebsch_Ishida_PRL_2007}%
  \BibitemOpen
  \bibfield  {author} {\bibinfo {author} {\bibfnamefont {A.}~\bibnamefont
  {Liebsch}}\ and\ \bibinfo {author} {\bibfnamefont {H.}~\bibnamefont
  {Ishida}},\ }\href {\doibase 10.1103/PhysRevLett.98.216403} {\bibfield
  {journal} {\bibinfo  {journal} {Phys. Rev. Lett.}\ }\textbf {\bibinfo
  {volume} {98}},\ \bibinfo {pages} {216403} (\bibinfo {year}
  {2007})}\BibitemShut {NoStop}%
\bibitem [{\citenamefont {Gorelov}\ \emph {et~al.}(2010)\citenamefont
  {Gorelov}, \citenamefont {Karolak}, \citenamefont {Wehling}, \citenamefont
  {Lechermann}, \citenamefont {Lichtenstein},\ and\ \citenamefont
  {Pavarini}}]{Gorelov_PRL_2010}%
  \BibitemOpen
  \bibfield  {author} {\bibinfo {author} {\bibfnamefont {E.}~\bibnamefont
  {Gorelov}}, \bibinfo {author} {\bibfnamefont {M.}~\bibnamefont {Karolak}},
  \bibinfo {author} {\bibfnamefont {T.~O.}\ \bibnamefont {Wehling}}, \bibinfo
  {author} {\bibfnamefont {F.}~\bibnamefont {Lechermann}}, \bibinfo {author}
  {\bibfnamefont {A.~I.}\ \bibnamefont {Lichtenstein}}, \ and\ \bibinfo
  {author} {\bibfnamefont {E.}~\bibnamefont {Pavarini}},\ }\href {\doibase
  10.1103/PhysRevLett.104.226401} {\bibfield  {journal} {\bibinfo  {journal}
  {Phys. Rev. Lett.}\ }\textbf {\bibinfo {volume} {104}},\ \bibinfo {pages}
  {226401} (\bibinfo {year} {2010})}\BibitemShut {NoStop}%
\bibitem [{\citenamefont {Zhang}\ and\ \citenamefont
  {Pavarini}(2017)}]{Zhang_PRB_2017}%
  \BibitemOpen
  \bibfield  {author} {\bibinfo {author} {\bibfnamefont {G.}~\bibnamefont
  {Zhang}}\ and\ \bibinfo {author} {\bibfnamefont {E.}~\bibnamefont
  {Pavarini}},\ }\href {\doibase 10.1103/PhysRevB.95.075145} {\bibfield
  {journal} {\bibinfo  {journal} {Phys. Rev. B}\ }\textbf {\bibinfo {volume}
  {95}},\ \bibinfo {pages} {075145} (\bibinfo {year} {2017})}\BibitemShut
  {NoStop}%
\bibitem [{\citenamefont {Han}\ and\ \citenamefont
  {Millis}(2018)}]{Han_Millis_PRL_2018}%
  \BibitemOpen
  \bibfield  {author} {\bibinfo {author} {\bibfnamefont {Q.}~\bibnamefont
  {Han}}\ and\ \bibinfo {author} {\bibfnamefont {A.}~\bibnamefont {Millis}},\
  }\href {\doibase 10.1103/PhysRevLett.121.067601} {\bibfield  {journal}
  {\bibinfo  {journal} {Phys. Rev. Lett.}\ }\textbf {\bibinfo {volume} {121}},\
  \bibinfo {pages} {067601} (\bibinfo {year} {2018})}\BibitemShut {NoStop}%
\bibitem [{\citenamefont {Kresse}\ and\ \citenamefont
  {Hafner}(1993)}]{Kresse_PRB_1993}%
  \BibitemOpen
  \bibfield  {author} {\bibinfo {author} {\bibfnamefont {G.}~\bibnamefont
  {Kresse}}\ and\ \bibinfo {author} {\bibfnamefont {J.}~\bibnamefont
  {Hafner}},\ }\href {\doibase 10.1103/PhysRevB.47.558} {\bibfield  {journal}
  {\bibinfo  {journal} {Phys. Rev. B}\ }\textbf {\bibinfo {volume} {47}},\
  \bibinfo {pages} {558} (\bibinfo {year} {1993})}\BibitemShut {NoStop}%
\bibitem [{\citenamefont {Kresse}\ and\ \citenamefont
  {Furthm\"uller}(1996{\natexlab{a}})}]{Kresse_CMS_1996}%
  \BibitemOpen
  \bibfield  {author} {\bibinfo {author} {\bibfnamefont {G.}~\bibnamefont
  {Kresse}}\ and\ \bibinfo {author} {\bibfnamefont {J.}~\bibnamefont
  {Furthm\"uller}},\ }\href {\doibase 10.1016/0927-0256(96)00008-0} {\bibfield
  {journal} {\bibinfo  {journal} {Comput. Mat. Sci.}\ }\textbf {\bibinfo
  {volume} {6}},\ \bibinfo {pages} {15} (\bibinfo {year}
  {1996}{\natexlab{a}})}\BibitemShut {NoStop}%
\bibitem [{\citenamefont {Kresse}\ and\ \citenamefont
  {Furthm\"uller}(1996{\natexlab{b}})}]{Kresse_PRB_1996}%
  \BibitemOpen
  \bibfield  {author} {\bibinfo {author} {\bibfnamefont {G.}~\bibnamefont
  {Kresse}}\ and\ \bibinfo {author} {\bibfnamefont {J.}~\bibnamefont
  {Furthm\"uller}},\ }\href {\doibase 10.1103/PhysRevB.54.11169} {\bibfield
  {journal} {\bibinfo  {journal} {Phys. Rev. B}\ }\textbf {\bibinfo {volume}
  {54}},\ \bibinfo {pages} {11169} (\bibinfo {year}
  {1996}{\natexlab{b}})}\BibitemShut {NoStop}%
\bibitem [{\citenamefont {Kresse}\ and\ \citenamefont
  {Joubert}(1999)}]{Kresse_PRB_1999}%
  \BibitemOpen
  \bibfield  {author} {\bibinfo {author} {\bibfnamefont {G.}~\bibnamefont
  {Kresse}}\ and\ \bibinfo {author} {\bibfnamefont {D.}~\bibnamefont
  {Joubert}},\ }\href {\doibase 10.1103/PhysRevB.59.1758} {\bibfield  {journal}
  {\bibinfo  {journal} {Phys. Rev. B}\ }\textbf {\bibinfo {volume} {59}},\
  \bibinfo {pages} {1758} (\bibinfo {year} {1999})}\BibitemShut {NoStop}%
\bibitem [{\citenamefont {Marzari}\ and\ \citenamefont
  {Vanderbilt}(1997)}]{Marzari_Vanderbilt_PRB_1997}%
  \BibitemOpen
  \bibfield  {author} {\bibinfo {author} {\bibfnamefont {N.}~\bibnamefont
  {Marzari}}\ and\ \bibinfo {author} {\bibfnamefont {D.}~\bibnamefont
  {Vanderbilt}},\ }\href {\doibase 10.1103/PhysRevB.56.12847} {\bibfield
  {journal} {\bibinfo  {journal} {Phys. Rev. B}\ }\textbf {\bibinfo {volume}
  {56}},\ \bibinfo {pages} {12847} (\bibinfo {year} {1997})}\BibitemShut
  {NoStop}%
\bibitem [{\citenamefont {Souza}\ \emph {et~al.}(2001)\citenamefont {Souza},
  \citenamefont {Marzari},\ and\ \citenamefont
  {Vanderbilt}}]{Souza_Marzari_PRB_2001}%
  \BibitemOpen
  \bibfield  {author} {\bibinfo {author} {\bibfnamefont {I.}~\bibnamefont
  {Souza}}, \bibinfo {author} {\bibfnamefont {N.}~\bibnamefont {Marzari}}, \
  and\ \bibinfo {author} {\bibfnamefont {D.}~\bibnamefont {Vanderbilt}},\
  }\href {\doibase 10.1103/PhysRevB.65.035109} {\bibfield  {journal} {\bibinfo
  {journal} {Phys. Rev. B}\ }\textbf {\bibinfo {volume} {65}},\ \bibinfo
  {pages} {035109} (\bibinfo {year} {2001})}\BibitemShut {NoStop}%
\bibitem [{\citenamefont {Dang}\ and\ \citenamefont
  {Millis}(2013)}]{Dang_Millis_PRB_2013}%
  \BibitemOpen
  \bibfield  {author} {\bibinfo {author} {\bibfnamefont {H.~T.}\ \bibnamefont
  {Dang}}\ and\ \bibinfo {author} {\bibfnamefont {A.~J.}\ \bibnamefont
  {Millis}},\ }\href {\doibase 10.1103/PhysRevB.87.155127} {\bibfield
  {journal} {\bibinfo  {journal} {Phys. Rev. B}\ }\textbf {\bibinfo {volume}
  {87}},\ \bibinfo {pages} {155127} (\bibinfo {year} {2013})}\BibitemShut
  {NoStop}%
\bibitem [{\citenamefont {Dang}\ \emph {et~al.}(2015)\citenamefont {Dang},
  \citenamefont {Mravlje}, \citenamefont {Georges},\ and\ \citenamefont
  {Millis}}]{Dang_Millis_PRB_2015}%
  \BibitemOpen
  \bibfield  {author} {\bibinfo {author} {\bibfnamefont {H.~T.}\ \bibnamefont
  {Dang}}, \bibinfo {author} {\bibfnamefont {J.}~\bibnamefont {Mravlje}},
  \bibinfo {author} {\bibfnamefont {A.}~\bibnamefont {Georges}}, \ and\
  \bibinfo {author} {\bibfnamefont {A.~J.}\ \bibnamefont {Millis}},\ }\href
  {\doibase 10.1103/PhysRevB.91.195149} {\bibfield  {journal} {\bibinfo
  {journal} {Phys. Rev. B}\ }\textbf {\bibinfo {volume} {91}},\ \bibinfo
  {pages} {195149} (\bibinfo {year} {2015})}\BibitemShut {NoStop}%
\bibitem [{\citenamefont {Han}\ \emph {et~al.}(2016)\citenamefont {Han},
  \citenamefont {Dang},\ and\ \citenamefont {Millis}}]{Han_Millis_PRB_2016}%
  \BibitemOpen
  \bibfield  {author} {\bibinfo {author} {\bibfnamefont {Q.}~\bibnamefont
  {Han}}, \bibinfo {author} {\bibfnamefont {H.~T.}\ \bibnamefont {Dang}}, \
  and\ \bibinfo {author} {\bibfnamefont {A.~J.}\ \bibnamefont {Millis}},\
  }\href {\doibase 10.1103/PhysRevB.93.155103} {\bibfield  {journal} {\bibinfo
  {journal} {Phys. Rev. B}\ }\textbf {\bibinfo {volume} {93}},\ \bibinfo
  {pages} {155103} (\bibinfo {year} {2016})}\BibitemShut {NoStop}%
\bibitem [{\citenamefont {Mravlje}\ \emph {et~al.}(2011)\citenamefont
  {Mravlje}, \citenamefont {Aichhorn}, \citenamefont {Miyake}, \citenamefont
  {Haule}, \citenamefont {Kotliar},\ and\ \citenamefont
  {Georges}}]{Mravlje2011}%
  \BibitemOpen
  \bibfield  {author} {\bibinfo {author} {\bibfnamefont {J.}~\bibnamefont
  {Mravlje}}, \bibinfo {author} {\bibfnamefont {M.}~\bibnamefont {Aichhorn}},
  \bibinfo {author} {\bibfnamefont {T.}~\bibnamefont {Miyake}}, \bibinfo
  {author} {\bibfnamefont {K.}~\bibnamefont {Haule}}, \bibinfo {author}
  {\bibfnamefont {G.}~\bibnamefont {Kotliar}}, \ and\ \bibinfo {author}
  {\bibfnamefont {A.}~\bibnamefont {Georges}},\ }\href {\doibase
  10.1103/PhysRevLett.106.096401} {\bibfield  {journal} {\bibinfo  {journal}
  {Phys. Rev. Lett.}\ }\textbf {\bibinfo {volume} {106}},\ \bibinfo {pages}
  {096401} (\bibinfo {year} {2011})}\BibitemShut {NoStop}%
\bibitem [{\citenamefont {Sutter}\ \emph
  {et~al.}(2017{\natexlab{a}})\citenamefont {Sutter}, \citenamefont {Fatuzzo},
  \citenamefont {Moser}, \citenamefont {Kim}, \citenamefont {Fittipaldi},
  \citenamefont {Vecchione}, \citenamefont {Granata}, \citenamefont {Sassa},
  \citenamefont {Cossalter}, \citenamefont {Gatti}, \citenamefont {Grioni},
  \citenamefont {Ronnow}, \citenamefont {Plumb}, \citenamefont {Matt},
  \citenamefont {Shi}, \citenamefont {Hoesch}, \citenamefont {Kim},
  \citenamefont {Chang}, \citenamefont {Jeng}, \citenamefont {Jozwiak},
  \citenamefont {Bostwick}, \citenamefont {Rotenberg}, \citenamefont {Georges},
  \citenamefont {Neupert},\ and\ \citenamefont
  {Chang}}]{sutter_Ca2RuO4_natcomm_2017}%
  \BibitemOpen
  \bibfield  {author} {\bibinfo {author} {\bibfnamefont {D.}~\bibnamefont
  {Sutter}}, \bibinfo {author} {\bibfnamefont {C.~G.}\ \bibnamefont {Fatuzzo}},
  \bibinfo {author} {\bibfnamefont {S.}~\bibnamefont {Moser}}, \bibinfo
  {author} {\bibfnamefont {M.}~\bibnamefont {Kim}}, \bibinfo {author}
  {\bibfnamefont {R.}~\bibnamefont {Fittipaldi}}, \bibinfo {author}
  {\bibfnamefont {A.}~\bibnamefont {Vecchione}}, \bibinfo {author}
  {\bibfnamefont {V.}~\bibnamefont {Granata}}, \bibinfo {author} {\bibfnamefont
  {Y.}~\bibnamefont {Sassa}}, \bibinfo {author} {\bibfnamefont
  {F.}~\bibnamefont {Cossalter}}, \bibinfo {author} {\bibfnamefont
  {G.}~\bibnamefont {Gatti}}, \bibinfo {author} {\bibfnamefont
  {M.}~\bibnamefont {Grioni}}, \bibinfo {author} {\bibfnamefont {H.~M.}\
  \bibnamefont {Ronnow}}, \bibinfo {author} {\bibfnamefont {N.~C.}\
  \bibnamefont {Plumb}}, \bibinfo {author} {\bibfnamefont {C.~E.}\ \bibnamefont
  {Matt}}, \bibinfo {author} {\bibfnamefont {M.}~\bibnamefont {Shi}}, \bibinfo
  {author} {\bibfnamefont {M.}~\bibnamefont {Hoesch}}, \bibinfo {author}
  {\bibfnamefont {T.~K.}\ \bibnamefont {Kim}}, \bibinfo {author} {\bibfnamefont
  {T.-R.}\ \bibnamefont {Chang}}, \bibinfo {author} {\bibfnamefont {H.-T.}\
  \bibnamefont {Jeng}}, \bibinfo {author} {\bibfnamefont {C.}~\bibnamefont
  {Jozwiak}}, \bibinfo {author} {\bibfnamefont {A.}~\bibnamefont {Bostwick}},
  \bibinfo {author} {\bibfnamefont {E.}~\bibnamefont {Rotenberg}}, \bibinfo
  {author} {\bibfnamefont {A.}~\bibnamefont {Georges}}, \bibinfo {author}
  {\bibfnamefont {T.}~\bibnamefont {Neupert}}, \ and\ \bibinfo {author}
  {\bibfnamefont {J.}~\bibnamefont {Chang}},\ }\href@noop {} {\bibfield
  {journal} {\bibinfo  {journal} {Nature communications}\ }\textbf {\bibinfo
  {volume} {8}},\ \bibinfo {pages} {15176} (\bibinfo {year}
  {2017}{\natexlab{a}})}\BibitemShut {NoStop}%
\bibitem [{\citenamefont {Hao}\ \emph {et~al.}(2019)\citenamefont {Hao},
  \citenamefont {Rubenstein},\ and\ \citenamefont {Shi}}]{hao_2019_auxiliary}%
  \BibitemOpen
  \bibfield  {author} {\bibinfo {author} {\bibfnamefont {H.}~\bibnamefont
  {Hao}}, \bibinfo {author} {\bibfnamefont {B.~M.}\ \bibnamefont {Rubenstein}},
  \ and\ \bibinfo {author} {\bibfnamefont {H.}~\bibnamefont {Shi}},\ }\href
  {\doibase 10.1103/PhysRevB.99.235142} {\bibfield  {journal} {\bibinfo
  {journal} {Phys. Rev. B}\ }\textbf {\bibinfo {volume} {99}},\ \bibinfo
  {pages} {235142} (\bibinfo {year} {2019})}\BibitemShut {NoStop}%
\bibitem [{\citenamefont {Kanamori}(1963)}]{Kanamori_PTP_1963}%
  \BibitemOpen
  \bibfield  {author} {\bibinfo {author} {\bibfnamefont {J.}~\bibnamefont
  {Kanamori}},\ }\href {\doibase 10.1143/PTP.30.275} {\bibfield  {journal}
  {\bibinfo  {journal} {Prog. Theor. Phys.}\ }\textbf {\bibinfo {volume}
  {30}},\ \bibinfo {pages} {275} (\bibinfo {year} {1963})}\BibitemShut
  {NoStop}%
\bibitem [{\citenamefont {Qin}\ \emph {et~al.}(2016)\citenamefont {Qin},
  \citenamefont {Shi},\ and\ \citenamefont {Zhang}}]{PhysRevB.94.235119}%
  \BibitemOpen
  \bibfield  {author} {\bibinfo {author} {\bibfnamefont {M.}~\bibnamefont
  {Qin}}, \bibinfo {author} {\bibfnamefont {H.}~\bibnamefont {Shi}}, \ and\
  \bibinfo {author} {\bibfnamefont {S.}~\bibnamefont {Zhang}},\ }\href
  {\doibase 10.1103/PhysRevB.94.235119} {\bibfield  {journal} {\bibinfo
  {journal} {Phys. Rev. B}\ }\textbf {\bibinfo {volume} {94}},\ \bibinfo
  {pages} {235119} (\bibinfo {year} {2016})}\BibitemShut {NoStop}%
\bibitem [{\citenamefont {Parcollet}\ \emph {et~al.}(2015)\citenamefont
  {Parcollet}, \citenamefont {Ferrero}, \citenamefont {Ayral}, \citenamefont
  {Hafermann}, \citenamefont {Krivenko}, \citenamefont {Messio},\ and\
  \citenamefont {Seth}}]{TRIQS_CPC_2015}%
  \BibitemOpen
  \bibfield  {author} {\bibinfo {author} {\bibfnamefont {O.}~\bibnamefont
  {Parcollet}}, \bibinfo {author} {\bibfnamefont {M.}~\bibnamefont {Ferrero}},
  \bibinfo {author} {\bibfnamefont {T.}~\bibnamefont {Ayral}}, \bibinfo
  {author} {\bibfnamefont {H.}~\bibnamefont {Hafermann}}, \bibinfo {author}
  {\bibfnamefont {I.}~\bibnamefont {Krivenko}}, \bibinfo {author}
  {\bibfnamefont {L.}~\bibnamefont {Messio}}, \ and\ \bibinfo {author}
  {\bibfnamefont {P.}~\bibnamefont {Seth}},\ }\href {\doibase
  http://dx.doi.org/10.1016/j.cpc.2015.04.023} {\bibfield  {journal} {\bibinfo
  {journal} {Comput. Phys. Commun.}\ }\textbf {\bibinfo {volume} {196}},\
  \bibinfo {pages} {398 } (\bibinfo {year} {2015})}\BibitemShut {NoStop}%
\bibitem [{\citenamefont {Seth}\ \emph {et~al.}(2016)\citenamefont {Seth},
  \citenamefont {Krivenko}, \citenamefont {Ferrero},\ and\ \citenamefont
  {Parcollet}}]{TRIQS/CTHYB_CPC_2016}%
  \BibitemOpen
  \bibfield  {author} {\bibinfo {author} {\bibfnamefont {P.}~\bibnamefont
  {Seth}}, \bibinfo {author} {\bibfnamefont {I.}~\bibnamefont {Krivenko}},
  \bibinfo {author} {\bibfnamefont {M.}~\bibnamefont {Ferrero}}, \ and\
  \bibinfo {author} {\bibfnamefont {O.}~\bibnamefont {Parcollet}},\ }\href
  {\doibase http://dx.doi.org/10.1016/j.cpc.2015.10.023} {\bibfield  {journal}
  {\bibinfo  {journal} {Comput. Phys. Commun.}\ }\textbf {\bibinfo {volume}
  {200}},\ \bibinfo {pages} {274 } (\bibinfo {year} {2016})}\BibitemShut
  {NoStop}%
\bibitem [{\citenamefont {Lebre~Alireza}\ \emph {et~al.}(2007)\citenamefont
  {Lebre~Alireza}, \citenamefont {Barakat}, \citenamefont {Cumberlidge},
  \citenamefont {Lonzarich}, \citenamefont {Nakamura},\ and\ \citenamefont
  {Maeno}}]{Lebre_Maeno_JPSJ_2007}%
  \BibitemOpen
  \bibfield  {author} {\bibinfo {author} {\bibfnamefont {P.}~\bibnamefont
  {Lebre~Alireza}}, \bibinfo {author} {\bibfnamefont {S.}~\bibnamefont
  {Barakat}}, \bibinfo {author} {\bibfnamefont {A.-M.}\ \bibnamefont
  {Cumberlidge}}, \bibinfo {author} {\bibfnamefont {G.}~\bibnamefont
  {Lonzarich}}, \bibinfo {author} {\bibfnamefont {F.}~\bibnamefont {Nakamura}},
  \ and\ \bibinfo {author} {\bibfnamefont {Y.}~\bibnamefont {Maeno}},\ }\href
  {\doibase 10.1143/JPSJS.76SA.216} {\bibfield  {journal} {\bibinfo  {journal}
  {J. Phys. Soc. Jpn.}\ }\textbf {\bibinfo {volume} {76}},\ \bibinfo {pages}
  {216} (\bibinfo {year} {2007})}\BibitemShut {NoStop}%
\bibitem [{\citenamefont {Sutter}\ \emph
  {et~al.}(2017{\natexlab{b}})\citenamefont {Sutter}, \citenamefont {Fatuzzo},
  \citenamefont {Moser}, \citenamefont {Kim}, \citenamefont {Fittipaldi},
  \citenamefont {Vecchione}, \citenamefont {Granata}, \citenamefont {Sassa},
  \citenamefont {Cossalter}, \citenamefont {Gatti}, \citenamefont {Grioni},
  \citenamefont {R{\o}znnow}, \citenamefont {Plumb}, \citenamefont {Matt},
  \citenamefont {Shi}, \citenamefont {Hoesch}, \citenamefont {Kim},
  \citenamefont {Chang}, \citenamefont {Jeng}, \citenamefont {Jozwiak},
  \citenamefont {Bostwick}, \citenamefont {Rotenberg}, \citenamefont {Georges},
  \citenamefont {Neupert},\ and\ \citenamefont {Chang}}]{Sutter_Chang_NC_2017}%
  \BibitemOpen
  \bibfield  {author} {\bibinfo {author} {\bibfnamefont {D.}~\bibnamefont
  {Sutter}}, \bibinfo {author} {\bibfnamefont {C.~G.}\ \bibnamefont {Fatuzzo}},
  \bibinfo {author} {\bibfnamefont {S.}~\bibnamefont {Moser}}, \bibinfo
  {author} {\bibfnamefont {M.}~\bibnamefont {Kim}}, \bibinfo {author}
  {\bibfnamefont {R.}~\bibnamefont {Fittipaldi}}, \bibinfo {author}
  {\bibfnamefont {A.}~\bibnamefont {Vecchione}}, \bibinfo {author}
  {\bibfnamefont {V.}~\bibnamefont {Granata}}, \bibinfo {author} {\bibfnamefont
  {Y.}~\bibnamefont {Sassa}}, \bibinfo {author} {\bibfnamefont
  {F.}~\bibnamefont {Cossalter}}, \bibinfo {author} {\bibfnamefont
  {G.}~\bibnamefont {Gatti}}, \bibinfo {author} {\bibfnamefont
  {M.}~\bibnamefont {Grioni}}, \bibinfo {author} {\bibfnamefont {H.~M.}\
  \bibnamefont {R{\o}znnow}}, \bibinfo {author} {\bibfnamefont {N.~C.}\
  \bibnamefont {Plumb}}, \bibinfo {author} {\bibfnamefont {C.~E.}\ \bibnamefont
  {Matt}}, \bibinfo {author} {\bibfnamefont {M.}~\bibnamefont {Shi}}, \bibinfo
  {author} {\bibfnamefont {M.}~\bibnamefont {Hoesch}}, \bibinfo {author}
  {\bibfnamefont {T.~K.}\ \bibnamefont {Kim}}, \bibinfo {author} {\bibfnamefont
  {T.-R.}\ \bibnamefont {Chang}}, \bibinfo {author} {\bibfnamefont {H.-T.}\
  \bibnamefont {Jeng}}, \bibinfo {author} {\bibfnamefont {C.}~\bibnamefont
  {Jozwiak}}, \bibinfo {author} {\bibfnamefont {A.}~\bibnamefont {Bostwick}},
  \bibinfo {author} {\bibfnamefont {E.}~\bibnamefont {Rotenberg}}, \bibinfo
  {author} {\bibfnamefont {A.}~\bibnamefont {Georges}}, \bibinfo {author}
  {\bibfnamefont {T.}~\bibnamefont {Neupert}}, \ and\ \bibinfo {author}
  {\bibfnamefont {J.}~\bibnamefont {Chang}},\ }\href
  {https://doi.org/10.1038/ncomms15176} {\bibfield  {journal} {\bibinfo
  {journal} {Nat. Commun.}\ }\textbf {\bibinfo {volume} {8}},\ \bibinfo {pages}
  {15176} (\bibinfo {year} {2017}{\natexlab{b}})}\BibitemShut {NoStop}%
\bibitem [{\citenamefont {Nakatsuji}\ and\ \citenamefont
  {Maeno}(2000)}]{Nakatsuji_PRL_2000}%
  \BibitemOpen
  \bibfield  {author} {\bibinfo {author} {\bibfnamefont {S.}~\bibnamefont
  {Nakatsuji}}\ and\ \bibinfo {author} {\bibfnamefont {Y.}~\bibnamefont
  {Maeno}},\ }\href {\doibase 10.1103/PhysRevLett.84.2666} {\bibfield
  {journal} {\bibinfo  {journal} {Phys. Rev. Lett.}\ }\textbf {\bibinfo
  {volume} {84}},\ \bibinfo {pages} {2666} (\bibinfo {year}
  {2000})}\BibitemShut {NoStop}%
\end{thebibliography}%
\end{document}